\documentclass[%
 reprint,
 amsmath,amssymb,
 aps,
prb,
floatfix,
]{revtex4-1}

\usepackage{graphicx}
\usepackage{dcolumn}
\usepackage{bm}

\usepackage{multirow}
\usepackage{array}

\begin{document}

\preprint{APS/123-QED}

\title{Controlling high-$Q$ trapped modes in polarization-insensitive all-dielectric metasurfaces}

\author{Andrey~Sayanskiy$^1$}
\author{Anton~S.~Kupriianov$^{2}$}  
\author{Su Xu$^{3,4}$}
\author{Polina~Kapitanova$^1$}%
\author{Victor~Dmitriev$^5$}
\author{Vyacheslav~V.~Khardikov$^{6,7}$}
\author{Vladimir~R.~Tuz$^{3,4,7}$}
 \email{tvr@jlu.edu.cn}
\affiliation{$^1$ITMO University, St.~Petersburg 197101, Russia}%
\affiliation{$^2$College of Physics, Jilin University, 2699 Qianjin St., Changchun 130012, China} 
\affiliation{$^3$State Key Laboratory of Integrated Optoelectronics, College of Electronic Science and Engineering, Jilin University, 2699 Qianjin St., Changchun 130012, China}
\affiliation{$^4$International Center of Future Science, Jilin University, 2699 Qianjin St., Changchun 130012, China}
\affiliation{$^5$Electrical Engineering Department, Federal University of Para, PO Box 8619, Agencia UFPA, CEP 66075-900 Belem, Para, Brazil}%
\affiliation{$^6$School of Radio Physics, V. N. Karazin Kharkiv National University, 4 Svobody Sq., Kharkiv 61022, Ukraine}
\affiliation{$^7$Institute of Radio Astronomy of National Academy of Sciences of Ukraine, 4 Mystetstv St., Kharkiv 61002, Ukraine}

\date{\today}

\begin{abstract}
We reveal peculiarities of the trapped (dark) mode excitation in a polarization-insensitive all-dielectric metasurface, whose unit super-cell is constructed by particularly arranging four cylindrical dielectric particles. Involving group-theoretical description we discuss in detail the effect of different orientations of particles within the super-cell on characteristics of the trapped mode. The theoretical predictions are confirmed by numerical simulations and experimental investigations. Since the metasurface is realized from simple dielectric particles without the use of any metallic components, they are feasibly scalable to both micro- and nanometer-size structures, and they can be employed in flat-optics platforms for realizing efficient light-matter interaction for multiple hotspot light localization, optical sensing, and highly-efficient light trapping.
\end{abstract}



\maketitle
\section{\label{intr}Introduction}
Today, with the advancements of nanotechnology, the science and engineering of \textit{subwavelength} light-matter interactions move to design and fabrication of nanostructures with desired optical properties.\cite{zheludev_science_2015} It allows one to construct optical components with previously unattainable characteristics. These also include \textit{planar} metamaterials (metasurfaces) made of metallic (plasmonic) particles which are extremely important parts of sensors, slow light and beam steering devices, holographic displays, near-IR tunable filters, fast optical inter-connectors, switchers, and amplitude modulators.

A particular class of novel metasurfaces based on dielectric nanoparticles has several significant advantages compared to plasmonic ones, especially concerning their low losses and fabrication techniques targeting to higher frequencies of operation.\cite{zheludev_naturemat_2012, kruk_acsphotonics_2017} They are so-called all-dielectric\cite{Zhao_MatToday_2009, jahani_NatNano_2016} metasurfaces composed of subwavelength dielectric particles made of high-refractive-index material.\cite{Baranov_Optica_2017} The particles are arranged into a lattice where each particle behaves as an individual resonator\cite{trubin2015lattices} sustaining a set of electric and magnetic dipolar and multipolar modes (referred to Mie-type modes\cite{Bohren_book}). An appearance of these modes can be spectrally controlled and engineered independently.\cite{Decker_AdvOptMat_2015}

Due to the unique features of optically induced electric and magnetic Mie-type modes of the dielectric particles, the all-dielectric metasurfaces are expected to complement or even replace different plasmonic components in a range of potential applications.\cite{Spinelli_NatComm_2012, Liu_OptExpress_2017, Bontempi_Nanoscale_2017} However, metasurfaces based on the Mie-type modes are still not sufficiently thin to compete with their metallic counterparts. An alternative approach is to access \textit{trapped} (dark) modes of the dielectric resonators, which allow deeper subwavelength thicknesses while still preserving a sharp resonant response.\cite{Khardikov_JOpt_2012, Zhang_OptExpress_2013, Campione_acsphotonics_2016, Tuz_OptExpress_2018} In the electromagnetic theory,\cite{Singh_PhysRevB_2009} the trapped modes are considered as some degenerate states that are not directly coupled to the field of incoming radiation, whereas they can be excited indirectly by removing the degeneracy. 

In particular, in order to lift the degeneracy and provide a necessary coupling of an incoming radiation with a trapped mode, some structural asymmetry can be introduced into the particles forming the metasurface\cite{Zouhdi_Advances_2003, Fedotov_PhysRevLett_2007} (recently, the trapped modes in such asymmetrical structures were referred to the phenomenon of bound states in the continuum (BIC) originated from distortion of the symmetry-protected bound state in the continuum\cite{Koshelev_PhysRevLett_2018}). As a side effect of the method, the resulting metasurface composed of asymmetrical particles becomes to be polarization sensitive even for a normally incident electromagnetic wave. It reduces the practical applicability of the metasurfaces based on trapped modes. In order to overcome this drawback, several designs of the metasurfaces composed of thin metallic particles have been proposed,\cite{Prosvirnin_ApplPhysLett_2009, Prosvirnin_JEMWA_2010, AlNaib_ApplPhysLett_2011, Tuz_EurPhys_2011, Tuz_JOpt_2012, Meng_MTT_2012, Tuong_ApplPhysLett_2013, Yu_JOpt_2013} whereas for the all-dielectric metasurfaces they are quite rare.\cite{ZhangFuli_OptExpress_2013, jain_advoptmater_2015, Kapitanova_AdvOptMat} It is due to a greater flexibility available in designs of thin metallic resonators. They can be given a rather complicate shape that provides a specific configuration of the \textit{surface} currents flow, which is impossible to realize in the volumetric dielectric resonators for the \textit{displacement} (polarization) currents. Therefore, in order to access a trapped mode in polarization-insensitive all-dielectric metasurfaces another approach should be applied.

As such a design of polarization-insensitive all-dielectric metasurfaces,\cite{jain_advoptmater_2015} a multi-layer configuration of dielectric resonators can be mentioned. In this configuration, in order to access a trapped mode, two additional scatterers of a particular form are attached diagonally to the upper and lower sides of the resonators forming metasurface. A multistep lift-off and deposition procedure is suggested for the resonators fabrication. While the metasurfaces consisting of ordinary (single-layer) asymmetric particles are fabricated and investigated,\cite{jain_advoptmater_2015} the structure composed of the multi-layer resonators has been investigated only numerically. Apparently this is due to the fact that the technology of their production, although feasible, is quite complex.

Alternatively, a trapped mode can be excited in the polarization-insensitive metasurfaces whose unit cell comprises several particles arranged specifically (so-called \textit{super-cell}\cite{Lim_OptExpress_2010, AlNaib_ApplPhysLett_2012, AlNaib_Conf_2014, Wu_OptExpress_2014, tuz_ACSPhotonics_2018}). 
Although these metasurfaces possess a more complicated unit cell, their production complexity is the same as for the metasurfaces based on the ordinary single-particle unit cells.

Following the concept of super-cell metasurfaces, in the present paper we employ group-theoretical predictions,\cite{Dmitriev_Metamat_2011, Dmitriev_IEEEAntennas_2013} numerical simulations and experimental study for both far-field and near-field characteristics to reveal conditions of the trapped mode excitation in a polarization-insensitive metasurface. We distinguish several super-cell's designs which provide efficient coupling of the all-dielectric metasurface with a linearly polarized incident wave via the trapped mode. They utilize particles with a short coaxial-sector notch made in a form of \textit{smile} since they can efficiently support the trapped mode.\cite{Tuz_OptExpress_2018}

\section{\label{mod}Theoretical description}

\begin{figure}[t!]
\centering
\includegraphics[width=1.0\linewidth]{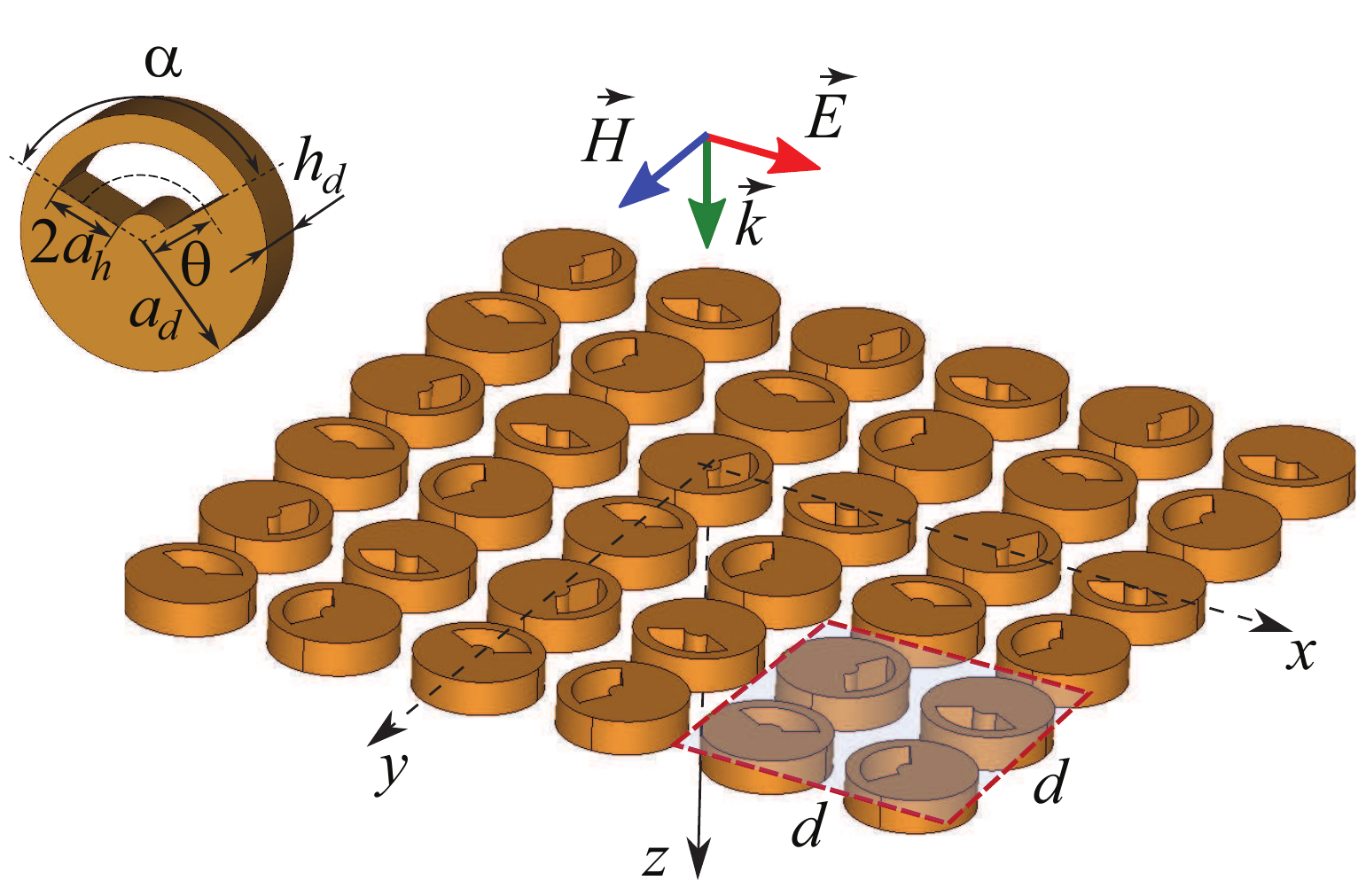}
\caption{Schematic view of an all-dielectric metasurface with the super-cell composed of four identical particles having coaxial-sector notch. $a_d$ and $h_d$ are the disks radius and height, respectively, $\theta$ is the radius of the sector mid-line, $2a_h$ is the notch width, and $\alpha$ is the sector opening angle. The disks are made from a nonmagnetic dielectric having permittivity $\varepsilon_d$. All disks are deposited equidistantly with period $d$. 
The lattice is buried into a dielectric host having permittivity $\varepsilon_s$ and thickness $h_s$ to form a whole metasurface (not shown in the figure).}
\label{fig:sketch}
\end{figure}

In order to ensure the completeness of our study, we start the discussion by demonstrating a polarization-sensitive response of the trapped mode excited by a linearly polarized wave in the  metasurface. 
We consider that the metasurface under study is illuminated by a normally incident ($\vec k = \{0,0,k_z\}$) linearly polarized wave with the electric field vector directed either along the $x$-axis ($\vec E = \{E_x,0,0\}$, $x$-polarized wave) or along the $y$-axis ($\vec E = \{0,E_y,0\}$, $y$-polarized wave) (see Fig.~\ref{fig:sketch}).

\begin{figure*}[htp]
\centering
\includegraphics[width=0.8\linewidth]{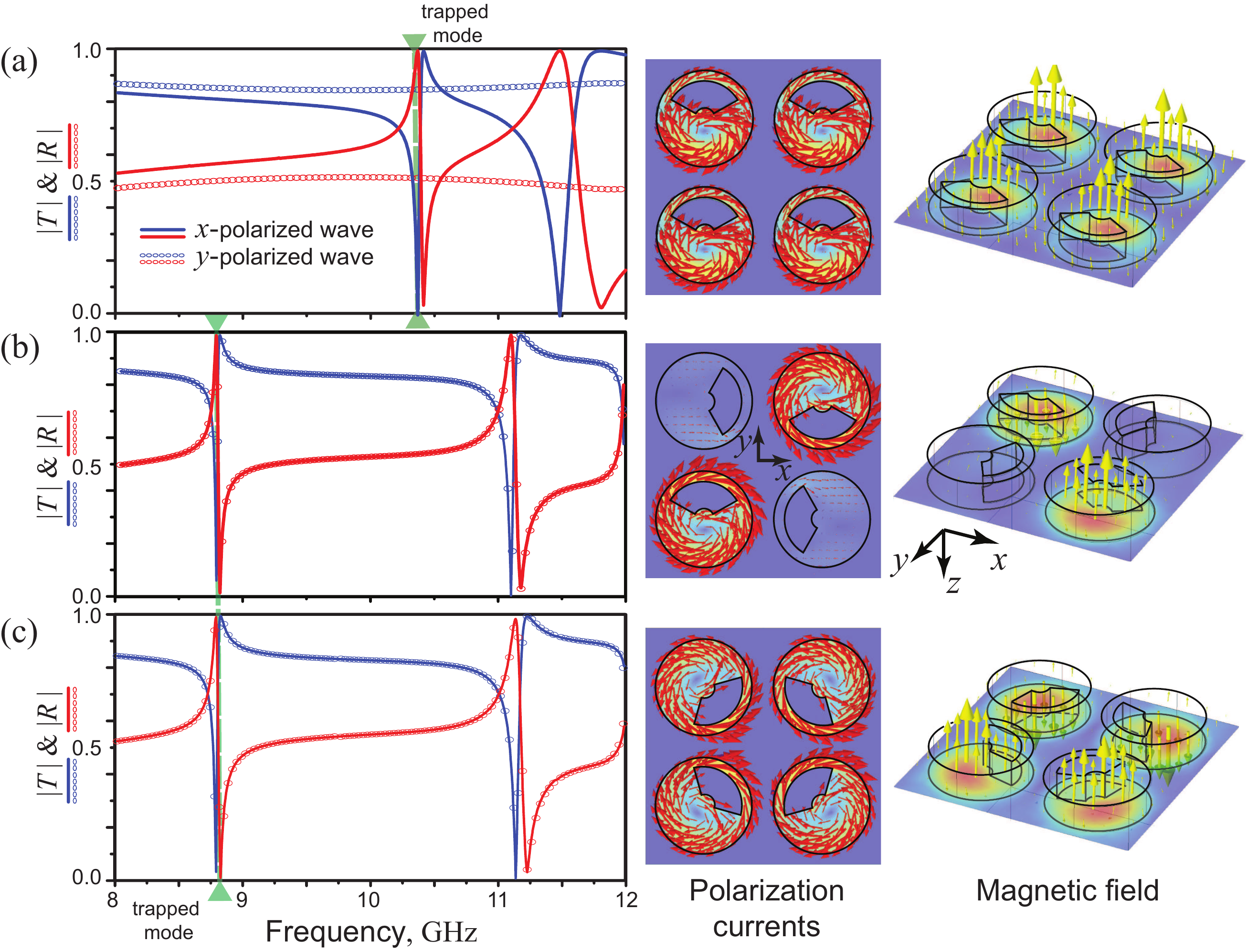}
\caption{Transmission ($T$) and reflection ($R$) coefficients of an all-dielectric (lossless) metasurface with the super-cell composed of particles oriented with symmetry (a) $C_s$, (b) $C_4$, and (c) $C_{4v}$. The geometrical parameters of the disk are  $a_d=4.5$ mm, $h_d=2.5$ mm, $\theta=2$ mm, $a_h=1$ mm, $\alpha=120^\circ$, the period of lattice is $d=22$ mm. The permittivities of disks and host are $\varepsilon_d=24$ and $\varepsilon_s=1.1$, respectively. The thickness of the host is $h_s = 20$ mm. The color maps demonstrate distributions of the polarization current (red arrows) and magnetic field (yellow arrows) calculated within the super-cell at the resonant frequency of trapped mode excited by the $x$-polarized incident wave.}
\label{fig:simulated}
\end{figure*} 

The numerical simulations of the electromagnetic response of the metasurface were performed with the use of RF module of commercial COMSOL Multiphysics\textsuperscript{\textregistered} finite-element electromagnetic solver. The Floquet-periodic boundary conditions were imposed on four sides of the unit cell to simulate the infinite two-dimensional array of resonators.\cite{comsol}

The point symmetry of the super-cell depends on the orientation of notches of the four disks. For the first design, symmetry of the super-cell in the $x-y$ plane is described by the group $C_s$ (in Sch\"oenflies notation\cite{barybin2002modern}) which contains only one vertical plane of symmetry $x = 0$. Under such a problem statement, the trapped mode appears to be excited only for the $x$-polarized wave (see Fig.~\ref{fig:simulated}(a) and the Supplemental Material\cite{Suppl_Mat}). In the spectra of the metasurface the trapped mode manifests itself as a peripheral lowest frequency (red-shifted\cite{Khardikov_JOpt_2012}) resonance (the corresponding resonance is distinguished on the spectral curves by the green arrows). The resonance acquires a sharp peak-and-trough (Fano) profile where extremes of transmission and reflection approach to 0 and 1 alternately, since in the simulation the losses in materials forming the metasurface are considered to be absent.

From the analysis of the polarization current and magnetic field distributions in the middle plane ($x-y$ plane at $z$ coordinate corresponding to the half height of the particles) it is revealed that the identified trapped mode resembles the characteristics of the lowest transverse electric (TE$_{01\delta}$) mode of the individual cylindrical dielectric resonator (see the color maps in Fig.~\ref{fig:simulated}(a) and the Supplemental Material\cite{Suppl_Mat}). At this state there is a specific distribution of the electromagnetic field within each particle, where the polarization currents have a circular flow twisting around the particle's center, whereas the magnetic field direction in the particle centers is oriented orthogonally to the metasurface plane forming out-of-plane magnetic dipole moment. All magnetic moments induced in the metasurface are oriented in the same direction demonstrating a dynamic ferromagnetic order.\cite{Wegener_PhysRevB_2009}

The most straightforward way to construct a polarization-insensitive metasurface based on the same particles is to rearrange them within the $2 \times 2$ super-cells, similarly to those proposed for the split ring based metasurfaces.\cite{AlNaib_ApplPhysLett_2012, AlNaib_Conf_2014, Wu_OptExpress_2014, Kapitanova_AdvOptMat} For the normal wave incidence, symmetry of the super-cell in the $x-y$ plane is described by the group $C_4$ which consists of the four-fold axis for rotation around the $z$-axis (Fig.~\ref{fig:simulated}(b)). It is mathematically proven that for the metasurfaces whose unit cell symmetry belongs to the rotational groups $C_n$, for $n>2$ there is polarization independence of the structure.\cite{Mackay_ElectronLett_1989} In Fig.~\ref{fig:simulated}(b) one can see that the spectral characteristics of the metasurface corresponding to excitation with two different polarizations are indeed identical.

For each polarization, the trapped mode is supported by a pair of active particles placed diagonally, while the remaining two resonators are inactive. In the color maps of Fig.~\ref{fig:simulated}(b) (Multimedia view) an active pair of particles is presented for the $x$-polarized wave. Another pair is active for the $y$-polarized wave and has the similar field distributions (are not presented here). Since only two particles are active within the super-cell, the lattice appears to be somewhat `sparse'. The out-of-plane magnetic moments induced in the active particles are oriented in the opposite directions resembling a dynamic antiferromagnetic order,\cite{Wegener_PhysRevB_2009, Decker_OptLett_2009} that leads to a shift of the resonance to the low frequency region in comparison with the $C_s$ design.


Next it is our goal is to find the polarization-insensitive configuration when all particles in the super-cell are active. For the metasurface under study there is a possibility to further increase the symmetry order of the super-cell via the design transition from the group $C_{4}$ to the group $C_{4v}$. The latter contains additionally four vertical planes of symmetry passing through the $z$-axis. They are $x=0$, $y=0$ and two diagonal planes. For this design the notches of the particles should be oriented either inward or outward the center of the super-cell. These two configurations are in fact identical, since the disks are situated equidistantly in the two-dimensional lattice, and thus there is arbitrariness in the super-cell choice. Remarkably, for such a design all four particles in the super-cell are active ones for the waves of both orthogonal polarizations.

\begin{figure}[!t]
\centering
\includegraphics[width=0.7\linewidth]{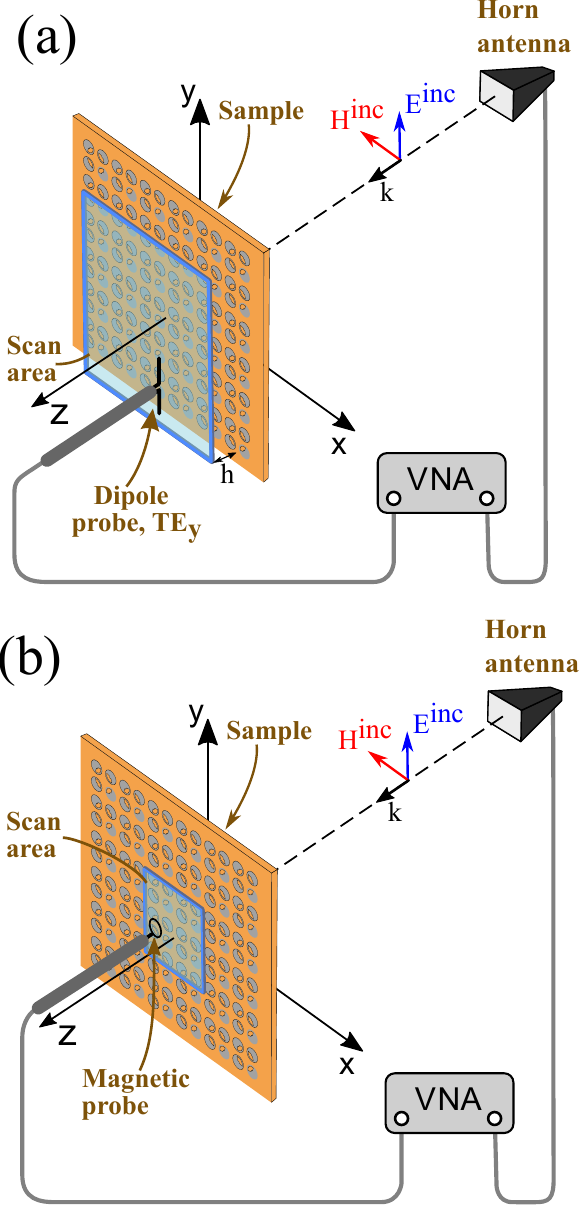}
\caption{Experimental setups for measurements of the (a) transmission and reflection spectra and (b) magnetic near-field distribution.}
\label{fig:setups}
\end{figure}

\begin{figure}[!t]
\centering
\includegraphics[width=1.0\linewidth]{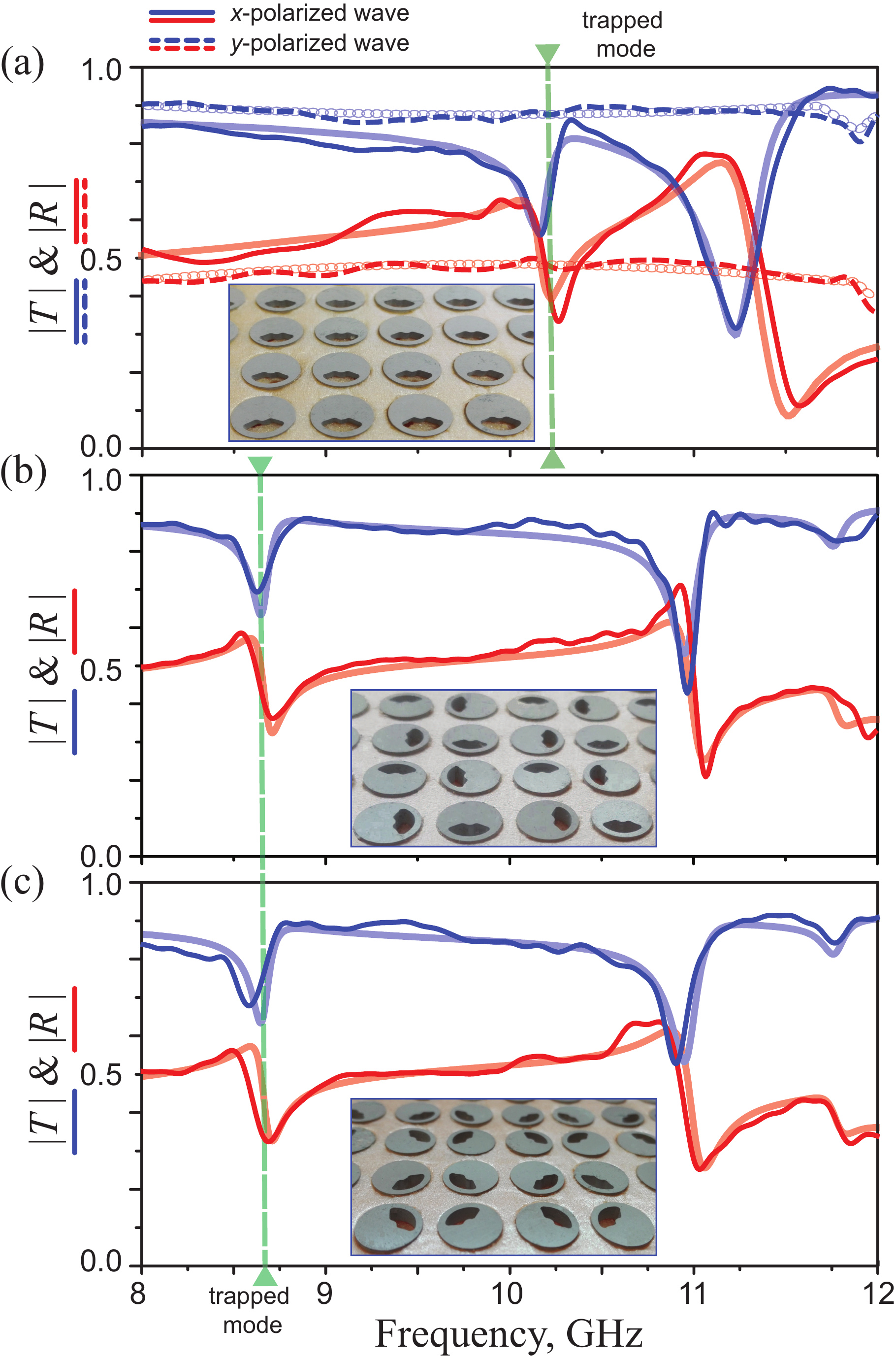}
\caption{Simulated (pale lines) and measured (bright lines) transmission and reflection coefficients of the all-dielectric metasurface with the super-cell composed of particles oriented with symmetry (a) $C_s$, (b) $C_4$, and (c) $C_{4v}$. In the simulation actual material losses ($\tan\delta=1 \times 10^{-3}$) in ceramic disks are taken into account, while the substrate is modeled as a lossless dielectric. The insets demonstrate fragments of the metasurface prototypes. The geometrical parameters of the metasurface are the same as in Fig.~\ref{fig:simulated}.}
\label{fig:measuredRT}
\end{figure}

The spectra of the metasurface whose unit cell possesses the higher symmetry and is described by the group $C_{4v}$ is presented in Fig.~\ref{fig:simulated}(c). It can be seen that in the frequency band of interest, the spectral curves for metasurfaces describing by the group $C_{4}$ and the group $C_{4v}$ are almost the same. The corresponding field distributions at the resonant frequency are presented in the color maps of Fig.~\ref{fig:simulated}(c) and the Supplemental Material.\cite{Suppl_Mat} Although in this design all particles are active, the dynamic antiferromagnetic order of the out-of-plane magnetic moments is preserved, and therefore the resonant frequency is very close to that of the $C_{4}$ design.

\section{\label{rep}Experimental verification}

In order to verify the polarization-insensitive appearance of a trapped mode, the prototypes of metasurface were fabricated and experimentally investigated in the microwave frequency range. As a dielectric material the Taizhou Wangling TP-series microwave ceramic characterized by the relative permittivity $\varepsilon_d=24$ and loss tangent $\tan \delta_d \le 1 \times 10^{-3}$ at 10 GHz has been used. The dielectric particles with the sizes mentioned in the caption of Fig.~\ref{fig:simulated} were fabricated with the use of precise mechanical cutting techniques. To arrange them, an array of holes was milled in a custom holder made of a Styrofoam material whose relative permittivity is $\varepsilon_s=1.1$ and thickness of the plate is $h_s = 20.0$~mm. The metasurface prototypes were constructed of $12 \times 12$ super-cells (so we used 576 particles in total) arranged in a lattice having the period $d=22$~mm.

At the first step the transmission and reflection spectra have been measured for all the metasurface prototypes. The common technique when the measurements are performed in the radiating near-field region and then transformed to the far-field zone has been used.\cite{johnson_IEEE_1973} During the investigation the prototype was fixed on the $2.0$~m distance from a rectangular linearly polarized broadband horn antenna as shown in Fig.~\ref{fig:setups}(a). The antenna generates a quasi-plane wave with required polarization. It is connected to the port of the Vector Network Analyzer (VNA) Agilent E8362C by a $50$~Ohm coaxial cable. An electrically small dipole probe oriented in parallel to the metasurface plane and connected to the second port of the VNA was used to detect the near electric field. During the measurements the probe was automatically moved in the $x-y$ plane over the scan area of $220 \times 220$~mm with a $5$~mm step at the distance of $30$~mm above the prototype surface. At each probe position both amplitude and phase of the transverse component of the electric field were sampled in the frequency range of interest ($8-12$~GHz). The same values of the electric field in the absence of the prototype were measured as a reference and the post-processing procedure was performed to obtain the transmission coefficient.\cite{johnson_IEEE_1973}  The similar procedure has been performed to measure the reflection coefficient excepting the scanning of the field above a metal plate placed instead of the metasurface prototype as a reference for the post-processing procedure. The measured transmission and reflection coefficients for all proposed designs are depicted in Fig.~\ref{fig:measuredRT} in comparison to the results of numerical simulation of actual metasurfaces. One can conclude that the measured data repeat well the simulated results. To not overload the picture in Figs.~\ref{fig:measuredRT}(b) and \ref{fig:measuredRT}(c) we presented curves only for the $x$-polarized wave, making sure that the spectra are identical for the waves of both orthogonal polarizations having carried out the corresponding measurements. 
 
\begin{figure}[!t]
\centering
\includegraphics[width=1.0\linewidth]{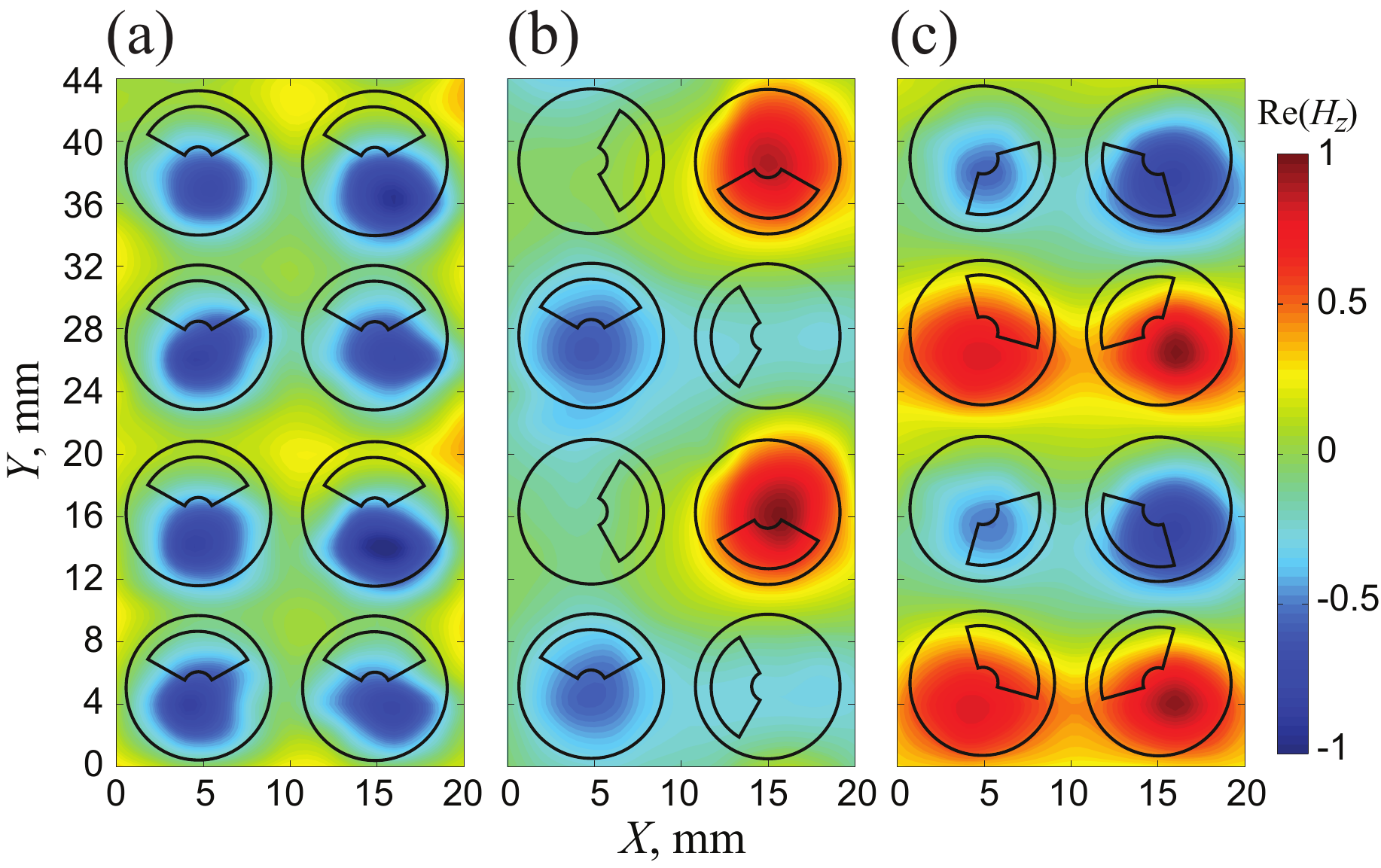}
\caption{Measured real part of the $H_z$ component of the magnetic near-field (out-of-plane magnetic moments) at the corresponding resonant frequency of trapped mode excited by the $x$-polarized incident wave in the all-dielectric metasurface with the super-cell composed of particles oriented with symmetry (a) $C_s$, (b) $C_4$, and (c) $C_{4v}$.}
\label{fig:measuredFields}
\end{figure} 

At the next step we have experimentally studied the near-field distribution of the metasurfaces. According to the results of our numerical simulations the electric field is mostly concentrated inside the particles, whereas the magnetic field is penetrating out of the resonators. Moreover, the magnetic moments are orthogonally oriented to the metasurface plane and demonstrate different patterns with respect to the particular resonator orderings. Thus, we have measured the magnetic near-field distribution for all the proposed designs. For that purpose we slightly changed the experimental setup. We used a small magnetic probe placed in the $x-y$ plane instead of the electric one (see Fig.~\ref{fig:setups} (b)). The normal component ($H_z$) of the magnetic near-field has been scanned over the area of $20 \times 44$~mm  with 1 mm step covering $1 \times 2$ super-cells. The near-field scanning was performed on the distance of 1 mm above the metasurface prototypes. The color maps of the measured near-field distribution confirming discussed above trapped mode resonant conditions and orientations of the out-of-plane magnetic moments are shown in Fig.~\ref{fig:measuredFields}.

\section{\label{concl}Conclusions}

In conclusion, we have studied the appearance of a trapped mode in several polarization-insensitive designs of the all-dielectric metasurface under the normal wave incidence condition.  The super-cell configuration for these designs is related to the symmetry groups $C_s$, $C_4$, and $C_{4v}$. Three different distributions of the magnetic moments are found to exist within the super-cell which resemble either ferromagnetic or antiferromagnetic order. An unexpected behaviour is that for the design related to the group $C_4$ only two particles are active in the super-cell, while other two are inactive. 

Although in our designs the particles with coaxial-sector notch in the shape of a smile are used, we argue the considered effects have a common nature and will also exist in particles with holes of a simpler shape, for example, in resonators with an off-centered round or rectangular hole that are easier to fabricate at the nanoscale. We believe that proposed designs can be useful in highly sensitive sensors, filters, and strong light matter interaction applications where polarization-insensitivity feature provides additional benefits.

\section*{\label{ack}Acknowledgments}
The experimental investigation of the reflection and transmission coefficients in the microwave frequency range has been supported by RSF (grant No. 17-19-01731). SX acknowledges support from the National Key Research and Development Program of China and National Natural Science Foundation of China (NSFC) under grant No. 61805097. VD thanks  Brazilian agency CNPq for  financial support. VRT acknowledges Jilin University's hospitality and financial support, and also thanks Yu. S. Kivshar for useful discussions and suggestions.

\bigskip

\bibliography{trapped_modes}

\begin{thebibliography}{45}%
\makeatletter
\providecommand \@ifxundefined [1]{%
 \@ifx{#1\undefined}
}%
\providecommand \@ifnum [1]{%
 \ifnum #1\expandafter \@firstoftwo
 \else \expandafter \@secondoftwo
 \fi
}%
\providecommand \@ifx [1]{%
 \ifx #1\expandafter \@firstoftwo
 \else \expandafter \@secondoftwo
 \fi
}%
\providecommand \natexlab [1]{#1}%
\providecommand \enquote  [1]{``#1''}%
\providecommand \bibnamefont  [1]{#1}%
\providecommand \bibfnamefont [1]{#1}%
\providecommand \citenamefont [1]{#1}%
\providecommand \href@noop [0]{\@secondoftwo}%
\providecommand \href [0]{\begingroup \@sanitize@url \@href}%
\providecommand \@href[1]{\@@startlink{#1}\@@href}%
\providecommand \@@href[1]{\endgroup#1\@@endlink}%
\providecommand \@sanitize@url [0]{\catcode `\\12\catcode `\$12\catcode
  `\&12\catcode `\#12\catcode `\^12\catcode `\_12\catcode `\%12\relax}%
\providecommand \@@startlink[1]{}%
\providecommand \@@endlink[0]{}%
\providecommand \url  [0]{\begingroup\@sanitize@url \@url }%
\providecommand \@url [1]{\endgroup\@href {#1}{\urlprefix }}%
\providecommand \urlprefix  [0]{URL }%
\providecommand \Eprint [0]{\href }%
\providecommand \doibase [0]{http://dx.doi.org/}%
\providecommand \selectlanguage [0]{\@gobble}%
\providecommand \bibinfo  [0]{\@secondoftwo}%
\providecommand \bibfield  [0]{\@secondoftwo}%
\providecommand \translation [1]{[#1]}%
\providecommand \BibitemOpen [0]{}%
\providecommand \bibitemStop [0]{}%
\providecommand \bibitemNoStop [0]{.\EOS\space}%
\providecommand \EOS [0]{\spacefactor3000\relax}%
\providecommand \BibitemShut  [1]{\csname bibitem#1\endcsname}%
\let\auto@bib@innerbib\@empty
\bibitem [{\citenamefont {Zheludev}(2015)}]{zheludev_science_2015}%
  \BibitemOpen
  \bibfield  {author} {\bibinfo {author} {\bibfnamefont {N.~I.}\ \bibnamefont
  {Zheludev}},\ }\href {\doibase 10.1126/science.aac4360} {\bibfield  {journal}
  {\bibinfo  {journal} {Science}\ }\textbf {\bibinfo {volume} {348}},\ \bibinfo
  {pages} {973} (\bibinfo {year} {2015})}\BibitemShut {NoStop}%
\bibitem [{\citenamefont {Zheludev}\ and\ \citenamefont
  {Kivshar}(2012)}]{zheludev_naturemat_2012}%
  \BibitemOpen
  \bibfield  {author} {\bibinfo {author} {\bibfnamefont {N.~I.}\ \bibnamefont
  {Zheludev}}\ and\ \bibinfo {author} {\bibfnamefont {Y.~S.}\ \bibnamefont
  {Kivshar}},\ }\href {\doibase 10.1038/nmat3431} {\bibfield  {journal}
  {\bibinfo  {journal} {Nat. Mater.}\ }\textbf {\bibinfo {volume} {11}},\
  \bibinfo {pages} {917–} (\bibinfo {year} {2012})}\BibitemShut {NoStop}%
\bibitem [{\citenamefont {Kruk}\ and\ \citenamefont
  {Kivshar}(2017)}]{kruk_acsphotonics_2017}%
  \BibitemOpen
  \bibfield  {author} {\bibinfo {author} {\bibfnamefont {S.}~\bibnamefont
  {Kruk}}\ and\ \bibinfo {author} {\bibfnamefont {Y.}~\bibnamefont {Kivshar}},\
  }\href {\doibase 10.1021/acsphotonics.7b01038} {\bibfield  {journal}
  {\bibinfo  {journal} {ACS Photonics}\ }\textbf {\bibinfo {volume} {4}},\
  \bibinfo {pages} {2638} (\bibinfo {year} {2017})}\BibitemShut {NoStop}%
\bibitem [{\citenamefont {Zhao}\ \emph {et~al.}(2009)\citenamefont {Zhao},
  \citenamefont {Zhou}, \citenamefont {Zhang},\ and\ \citenamefont
  {Lippens}}]{Zhao_MatToday_2009}%
  \BibitemOpen
  \bibfield  {author} {\bibinfo {author} {\bibfnamefont {Q.}~\bibnamefont
  {Zhao}}, \bibinfo {author} {\bibfnamefont {J.}~\bibnamefont {Zhou}}, \bibinfo
  {author} {\bibfnamefont {F.}~\bibnamefont {Zhang}}, \ and\ \bibinfo {author}
  {\bibfnamefont {D.}~\bibnamefont {Lippens}},\ }\href {\doibase
  https://doi.org/10.1016/S1369-7021(09)70318-9} {\bibfield  {journal}
  {\bibinfo  {journal} {Mater. Today}\ }\textbf {\bibinfo {volume} {12}},\
  \bibinfo {pages} {60} (\bibinfo {year} {2009})}\BibitemShut {NoStop}%
\bibitem [{\citenamefont {Jahani}\ and\ \citenamefont
  {Jacob}(2016)}]{jahani_NatNano_2016}%
  \BibitemOpen
  \bibfield  {author} {\bibinfo {author} {\bibfnamefont {S.}~\bibnamefont
  {Jahani}}\ and\ \bibinfo {author} {\bibfnamefont {Z.}~\bibnamefont {Jacob}},\
  }\href {\doibase 10.1038/nnano.2015.304} {\bibfield  {journal} {\bibinfo
  {journal} {Nat. Nanotechnol.}\ }\textbf {\bibinfo {volume} {11}},\ \bibinfo
  {pages} {23} (\bibinfo {year} {2016})}\BibitemShut {NoStop}%
\bibitem [{\citenamefont {Baranov}\ \emph {et~al.}(2017)\citenamefont
  {Baranov}, \citenamefont {Zuev}, \citenamefont {Lepeshov}, \citenamefont
  {Kotov}, \citenamefont {Krasnok}, \citenamefont {Evlyukhin},\ and\
  \citenamefont {Chichkov}}]{Baranov_Optica_2017}%
  \BibitemOpen
  \bibfield  {author} {\bibinfo {author} {\bibfnamefont {D.~G.}\ \bibnamefont
  {Baranov}}, \bibinfo {author} {\bibfnamefont {D.~A.}\ \bibnamefont {Zuev}},
  \bibinfo {author} {\bibfnamefont {S.~I.}\ \bibnamefont {Lepeshov}}, \bibinfo
  {author} {\bibfnamefont {O.~V.}\ \bibnamefont {Kotov}}, \bibinfo {author}
  {\bibfnamefont {A.~E.}\ \bibnamefont {Krasnok}}, \bibinfo {author}
  {\bibfnamefont {A.~B.}\ \bibnamefont {Evlyukhin}}, \ and\ \bibinfo {author}
  {\bibfnamefont {B.~N.}\ \bibnamefont {Chichkov}},\ }\href {\doibase
  10.1364/OPTICA.4.000814} {\bibfield  {journal} {\bibinfo  {journal} {Optica}\
  }\textbf {\bibinfo {volume} {4}},\ \bibinfo {pages} {814} (\bibinfo {year}
  {2017})}\BibitemShut {NoStop}%
\bibitem [{\citenamefont {Trubin}(2016)}]{trubin2015lattices}%
  \BibitemOpen
  \bibfield  {author} {\bibinfo {author} {\bibfnamefont {A.}~\bibnamefont
  {Trubin}},\ }\href {\doibase 10.1007/978-3-319-25148-6} {\emph {\bibinfo
  {title} {Lattices of Dielectric Resonators}}},\ \bibinfo {series} {Springer
  Series in Advanced Microelectronics}, Vol.~\bibinfo {volume} {53}\ (\bibinfo
  {publisher} {Springer Cham, Heidelberg},\ \bibinfo {year} {2016})\BibitemShut
  {NoStop}%
\bibitem [{\citenamefont {Bohren}\ and\ \citenamefont
  {Huffmann}(2010)}]{Bohren_book}%
  \BibitemOpen
  \bibfield  {author} {\bibinfo {author} {\bibfnamefont {C.~F.}\ \bibnamefont
  {Bohren}}\ and\ \bibinfo {author} {\bibfnamefont {D.~R.}\ \bibnamefont
  {Huffmann}},\ }\href@noop {} {\emph {\bibinfo {title} {Absorption and
  Scattering of Light by Small Particles}}}\ (\bibinfo  {publisher}
  {Wiley-Interscience, New York},\ \bibinfo {year} {2010})\BibitemShut
  {NoStop}%
\bibitem [{\citenamefont {Decker}\ \emph {et~al.}(2015)\citenamefont {Decker},
  \citenamefont {Staude}, \citenamefont {Falkner}, \citenamefont {Dominguez},
  \citenamefont {Neshev}, \citenamefont {Brener}, \citenamefont {Pertsch},\
  and\ \citenamefont {Kivshar}}]{Decker_AdvOptMat_2015}%
  \BibitemOpen
  \bibfield  {author} {\bibinfo {author} {\bibfnamefont {M.}~\bibnamefont
  {Decker}}, \bibinfo {author} {\bibfnamefont {I.}~\bibnamefont {Staude}},
  \bibinfo {author} {\bibfnamefont {M.}~\bibnamefont {Falkner}}, \bibinfo
  {author} {\bibfnamefont {J.}~\bibnamefont {Dominguez}}, \bibinfo {author}
  {\bibfnamefont {D.~N.}\ \bibnamefont {Neshev}}, \bibinfo {author}
  {\bibfnamefont {I.}~\bibnamefont {Brener}}, \bibinfo {author} {\bibfnamefont
  {T.}~\bibnamefont {Pertsch}}, \ and\ \bibinfo {author} {\bibfnamefont
  {Y.~S.}\ \bibnamefont {Kivshar}},\ }\href {\doibase 10.1002/adom.201400584}
  {\bibfield  {journal} {\bibinfo  {journal} {Adv. Opt. Mater.}\ }\textbf
  {\bibinfo {volume} {3}},\ \bibinfo {pages} {813–} (\bibinfo {year}
  {2015})}\BibitemShut {NoStop}%
\bibitem [{\citenamefont {Spinelli}\ \emph {et~al.}(2012)\citenamefont
  {Spinelli}, \citenamefont {Verschuuren},\ and\ \citenamefont
  {Polman}}]{Spinelli_NatComm_2012}%
  \BibitemOpen
  \bibfield  {author} {\bibinfo {author} {\bibfnamefont {P.}~\bibnamefont
  {Spinelli}}, \bibinfo {author} {\bibfnamefont {M.~A.}\ \bibnamefont
  {Verschuuren}}, \ and\ \bibinfo {author} {\bibfnamefont {A.}~\bibnamefont
  {Polman}},\ }\href {\doibase 10.1038/ncomms1691} {\bibfield  {journal}
  {\bibinfo  {journal} {Nat. Commun.}\ }\textbf {\bibinfo {volume} {3}},\
  \bibinfo {pages} {692} (\bibinfo {year} {2012})}\BibitemShut {NoStop}%
\bibitem [{\citenamefont {Liu}\ \emph {et~al.}(2017)\citenamefont {Liu},
  \citenamefont {Fan}, \citenamefont {Shadrivov},\ and\ \citenamefont
  {Padilla}}]{Liu_OptExpress_2017}%
  \BibitemOpen
  \bibfield  {author} {\bibinfo {author} {\bibfnamefont {X.}~\bibnamefont
  {Liu}}, \bibinfo {author} {\bibfnamefont {K.}~\bibnamefont {Fan}}, \bibinfo
  {author} {\bibfnamefont {I.~V.}\ \bibnamefont {Shadrivov}}, \ and\ \bibinfo
  {author} {\bibfnamefont {W.~J.}\ \bibnamefont {Padilla}},\ }\href {\doibase
  10.1364/OE.25.000191} {\bibfield  {journal} {\bibinfo  {journal} {Opt.
  Express}\ }\textbf {\bibinfo {volume} {25}},\ \bibinfo {pages} {191}
  (\bibinfo {year} {2017})}\BibitemShut {NoStop}%
\bibitem [{\citenamefont {Bontempi}\ \emph {et~al.}(2017)\citenamefont
  {Bontempi}, \citenamefont {Chong}, \citenamefont {Orton}, \citenamefont
  {Staude}, \citenamefont {Choi}, \citenamefont {Alessandri}, \citenamefont
  {Kivshar},\ and\ \citenamefont {Neshev}}]{Bontempi_Nanoscale_2017}%
  \BibitemOpen
  \bibfield  {author} {\bibinfo {author} {\bibfnamefont {N.}~\bibnamefont
  {Bontempi}}, \bibinfo {author} {\bibfnamefont {K.~E.}\ \bibnamefont {Chong}},
  \bibinfo {author} {\bibfnamefont {H.~W.}\ \bibnamefont {Orton}}, \bibinfo
  {author} {\bibfnamefont {I.}~\bibnamefont {Staude}}, \bibinfo {author}
  {\bibfnamefont {D.-Y.}\ \bibnamefont {Choi}}, \bibinfo {author}
  {\bibfnamefont {I.}~\bibnamefont {Alessandri}}, \bibinfo {author}
  {\bibfnamefont {Y.~S.}\ \bibnamefont {Kivshar}}, \ and\ \bibinfo {author}
  {\bibfnamefont {D.~N.}\ \bibnamefont {Neshev}},\ }\href {\doibase
  10.1039/C6NR07904K} {\bibfield  {journal} {\bibinfo  {journal} {Nanoscale}\
  }\textbf {\bibinfo {volume} {9}},\ \bibinfo {pages} {4972} (\bibinfo {year}
  {2017})}\BibitemShut {NoStop}%
\bibitem [{\citenamefont {Khardikov}\ \emph {et~al.}(2012)\citenamefont
  {Khardikov}, \citenamefont {Iarko},\ and\ \citenamefont
  {Prosvirnin}}]{Khardikov_JOpt_2012}%
  \BibitemOpen
  \bibfield  {author} {\bibinfo {author} {\bibfnamefont {V.~V.}\ \bibnamefont
  {Khardikov}}, \bibinfo {author} {\bibfnamefont {E.~O.}\ \bibnamefont
  {Iarko}}, \ and\ \bibinfo {author} {\bibfnamefont {S.~L.}\ \bibnamefont
  {Prosvirnin}},\ }\href {http://stacks.iop.org/2040-8986/14/i=3/a=035103}
  {\bibfield  {journal} {\bibinfo  {journal} {J. Opt.}\ }\textbf {\bibinfo
  {volume} {14}},\ \bibinfo {pages} {035103} (\bibinfo {year}
  {2012})}\BibitemShut {NoStop}%
\bibitem [{\citenamefont {Zhang}\ \emph
  {et~al.}(2013{\natexlab{a}})\citenamefont {Zhang}, \citenamefont
  {MacDonald},\ and\ \citenamefont {Zheludev}}]{Zhang_OptExpress_2013}%
  \BibitemOpen
  \bibfield  {author} {\bibinfo {author} {\bibfnamefont {J.}~\bibnamefont
  {Zhang}}, \bibinfo {author} {\bibfnamefont {K.~F.}\ \bibnamefont
  {MacDonald}}, \ and\ \bibinfo {author} {\bibfnamefont {N.~I.}\ \bibnamefont
  {Zheludev}},\ }\href {\doibase 10.1364/OE.21.026721} {\bibfield  {journal}
  {\bibinfo  {journal} {Opt. Express}\ }\textbf {\bibinfo {volume} {21}},\
  \bibinfo {pages} {26721} (\bibinfo {year} {2013}{\natexlab{a}})}\BibitemShut
  {NoStop}%
\bibitem [{\citenamefont {Campione}\ \emph {et~al.}(2016)\citenamefont
  {Campione}, \citenamefont {Liu}, \citenamefont {Basilio}, \citenamefont
  {Warne}, \citenamefont {Langston}, \citenamefont {Luk}, \citenamefont
  {Wendt}, \citenamefont {Reno}, \citenamefont {Keeler}, \citenamefont
  {Brener},\ and\ \citenamefont {Sinclair}}]{Campione_acsphotonics_2016}%
  \BibitemOpen
  \bibfield  {author} {\bibinfo {author} {\bibfnamefont {S.}~\bibnamefont
  {Campione}}, \bibinfo {author} {\bibfnamefont {S.}~\bibnamefont {Liu}},
  \bibinfo {author} {\bibfnamefont {L.~I.}\ \bibnamefont {Basilio}}, \bibinfo
  {author} {\bibfnamefont {L.~K.}\ \bibnamefont {Warne}}, \bibinfo {author}
  {\bibfnamefont {W.~L.}\ \bibnamefont {Langston}}, \bibinfo {author}
  {\bibfnamefont {T.~S.}\ \bibnamefont {Luk}}, \bibinfo {author} {\bibfnamefont
  {J.~R.}\ \bibnamefont {Wendt}}, \bibinfo {author} {\bibfnamefont {J.~L.}\
  \bibnamefont {Reno}}, \bibinfo {author} {\bibfnamefont {G.~A.}\ \bibnamefont
  {Keeler}}, \bibinfo {author} {\bibfnamefont {I.}~\bibnamefont {Brener}}, \
  and\ \bibinfo {author} {\bibfnamefont {M.~B.}\ \bibnamefont {Sinclair}},\
  }\href {\doibase 10.1021/acsphotonics.6b00556} {\bibfield  {journal}
  {\bibinfo  {journal} {ACS Photonics}\ }\textbf {\bibinfo {volume} {3}},\
  \bibinfo {pages} {2362} (\bibinfo {year} {2016})}\BibitemShut {NoStop}%
\bibitem [{\citenamefont {Tuz}\ \emph {et~al.}(2018{\natexlab{a}})\citenamefont
  {Tuz}, \citenamefont {Khardikov}, \citenamefont {Kupriianov}, \citenamefont
  {Domina}, \citenamefont {Xu}, \citenamefont {Wang},\ and\ \citenamefont
  {Sun}}]{Tuz_OptExpress_2018}%
  \BibitemOpen
  \bibfield  {author} {\bibinfo {author} {\bibfnamefont {V.~R.}\ \bibnamefont
  {Tuz}}, \bibinfo {author} {\bibfnamefont {V.~V.}\ \bibnamefont {Khardikov}},
  \bibinfo {author} {\bibfnamefont {A.~S.}\ \bibnamefont {Kupriianov}},
  \bibinfo {author} {\bibfnamefont {K.~L.}\ \bibnamefont {Domina}}, \bibinfo
  {author} {\bibfnamefont {S.}~\bibnamefont {Xu}}, \bibinfo {author}
  {\bibfnamefont {H.}~\bibnamefont {Wang}}, \ and\ \bibinfo {author}
  {\bibfnamefont {H.-B.}\ \bibnamefont {Sun}},\ }\href {\doibase
  10.1364/OE.26.002905} {\bibfield  {journal} {\bibinfo  {journal} {Opt.
  Express}\ }\textbf {\bibinfo {volume} {26}},\ \bibinfo {pages} {2905}
  (\bibinfo {year} {2018}{\natexlab{a}})}\BibitemShut {NoStop}%
\bibitem [{\citenamefont {Singh}\ \emph {et~al.}(2009)\citenamefont {Singh},
  \citenamefont {Rockstuhl}, \citenamefont {Lederer},\ and\ \citenamefont
  {Zhang}}]{Singh_PhysRevB_2009}%
  \BibitemOpen
  \bibfield  {author} {\bibinfo {author} {\bibfnamefont {R.}~\bibnamefont
  {Singh}}, \bibinfo {author} {\bibfnamefont {C.}~\bibnamefont {Rockstuhl}},
  \bibinfo {author} {\bibfnamefont {F.}~\bibnamefont {Lederer}}, \ and\
  \bibinfo {author} {\bibfnamefont {W.}~\bibnamefont {Zhang}},\ }\href
  {\doibase 10.1103/PhysRevB.79.085111} {\bibfield  {journal} {\bibinfo
  {journal} {Phys. Rev. B}\ }\textbf {\bibinfo {volume} {79}},\ \bibinfo
  {pages} {085111} (\bibinfo {year} {2009})}\BibitemShut {NoStop}%
\bibitem [{\citenamefont {Prosvirnin}\ and\ \citenamefont
  {Zouhdi}(2003)}]{Zouhdi_Advances_2003}%
  \BibitemOpen
  \bibfield  {author} {\bibinfo {author} {\bibfnamefont {S.}~\bibnamefont
  {Prosvirnin}}\ and\ \bibinfo {author} {\bibfnamefont {S.}~\bibnamefont
  {Zouhdi}},\ }in\ \href@noop {} {\emph {\bibinfo {booktitle} {Advances in
  Electromagnetics of Complex Media and Metamaterials}}},\ \bibinfo {editor}
  {edited by\ \bibinfo {editor} {\bibfnamefont {S.}~\bibnamefont {Zouhdi}}\
  and\ \bibinfo {editor} {\bibfnamefont {M.}~\bibnamefont {Arsalane}}}\
  (\bibinfo  {publisher} {Kluwer Academic Publishers},\ \bibinfo {address} {the
  Netherlands},\ \bibinfo {year} {2003})\ pp.\ \bibinfo {pages}
  {281--290}\BibitemShut {NoStop}%
\bibitem [{\citenamefont {Fedotov}\ \emph {et~al.}(2007)\citenamefont
  {Fedotov}, \citenamefont {Rose}, \citenamefont {Prosvirnin}, \citenamefont
  {Papasimakis},\ and\ \citenamefont {Zheludev}}]{Fedotov_PhysRevLett_2007}%
  \BibitemOpen
  \bibfield  {author} {\bibinfo {author} {\bibfnamefont {V.~A.}\ \bibnamefont
  {Fedotov}}, \bibinfo {author} {\bibfnamefont {M.}~\bibnamefont {Rose}},
  \bibinfo {author} {\bibfnamefont {S.~L.}\ \bibnamefont {Prosvirnin}},
  \bibinfo {author} {\bibfnamefont {N.}~\bibnamefont {Papasimakis}}, \ and\
  \bibinfo {author} {\bibfnamefont {N.~I.}\ \bibnamefont {Zheludev}},\ }\href
  {\doibase 10.1103/PhysRevLett.99.147401} {\bibfield  {journal} {\bibinfo
  {journal} {Phys. Rev. Lett.}\ }\textbf {\bibinfo {volume} {99}},\ \bibinfo
  {pages} {147401} (\bibinfo {year} {2007})}\BibitemShut {NoStop}%
\bibitem [{\citenamefont {Koshelev}\ \emph {et~al.}(2018)\citenamefont
  {Koshelev}, \citenamefont {Lepeshov}, \citenamefont {Liu}, \citenamefont
  {Bogdanov},\ and\ \citenamefont {Kivshar}}]{Koshelev_PhysRevLett_2018}%
  \BibitemOpen
  \bibfield  {author} {\bibinfo {author} {\bibfnamefont {K.}~\bibnamefont
  {Koshelev}}, \bibinfo {author} {\bibfnamefont {S.}~\bibnamefont {Lepeshov}},
  \bibinfo {author} {\bibfnamefont {M.}~\bibnamefont {Liu}}, \bibinfo {author}
  {\bibfnamefont {A.}~\bibnamefont {Bogdanov}}, \ and\ \bibinfo {author}
  {\bibfnamefont {Y.}~\bibnamefont {Kivshar}},\ }\href {\doibase
  10.1103/PhysRevLett.121.193903} {\bibfield  {journal} {\bibinfo  {journal}
  {Phys. Rev. Lett.}\ }\textbf {\bibinfo {volume} {121}},\ \bibinfo {pages}
  {193903} (\bibinfo {year} {2018})}\BibitemShut {NoStop}%
\bibitem [{\citenamefont {Papasimakis}\ \emph {et~al.}(2009)\citenamefont
  {Papasimakis}, \citenamefont {Fu}, \citenamefont {Fedotov}, \citenamefont
  {Prosvirnin}, \citenamefont {Tsai},\ and\ \citenamefont
  {Zheludev}}]{Prosvirnin_ApplPhysLett_2009}%
  \BibitemOpen
  \bibfield  {author} {\bibinfo {author} {\bibfnamefont {N.}~\bibnamefont
  {Papasimakis}}, \bibinfo {author} {\bibfnamefont {Y.~H.}\ \bibnamefont {Fu}},
  \bibinfo {author} {\bibfnamefont {V.~A.}\ \bibnamefont {Fedotov}}, \bibinfo
  {author} {\bibfnamefont {S.~L.}\ \bibnamefont {Prosvirnin}}, \bibinfo
  {author} {\bibfnamefont {D.~P.}\ \bibnamefont {Tsai}}, \ and\ \bibinfo
  {author} {\bibfnamefont {N.~I.}\ \bibnamefont {Zheludev}},\ }\href {\doibase
  10.1063/1.3138868} {\bibfield  {journal} {\bibinfo  {journal} {Appl. Phys.
  Lett.}\ }\textbf {\bibinfo {volume} {94}},\ \bibinfo {pages} {211902}
  (\bibinfo {year} {2009})}\BibitemShut {NoStop}%
\bibitem [{\citenamefont {Kawakatsu}\ \emph {et~al.}(2010)\citenamefont
  {Kawakatsu}, \citenamefont {Dmitriev},\ and\ \citenamefont
  {Prosvirnin}}]{Prosvirnin_JEMWA_2010}%
  \BibitemOpen
  \bibfield  {author} {\bibinfo {author} {\bibfnamefont {M.~N.}\ \bibnamefont
  {Kawakatsu}}, \bibinfo {author} {\bibfnamefont {V.~A.}\ \bibnamefont
  {Dmitriev}}, \ and\ \bibinfo {author} {\bibfnamefont {S.~L.}\ \bibnamefont
  {Prosvirnin}},\ }\href {\doibase 10.1163/156939310790735741} {\bibfield
  {journal} {\bibinfo  {journal} {J. Electromagn. Waves Appl.}\ }\textbf
  {\bibinfo {volume} {24}},\ \bibinfo {pages} {261} (\bibinfo {year}
  {2010})}\BibitemShut {NoStop}%
\bibitem [{\citenamefont {Al-Naib}\ \emph {et~al.}(2011)\citenamefont
  {Al-Naib}, \citenamefont {Jansen}, \citenamefont {Born},\ and\ \citenamefont
  {Koch}}]{AlNaib_ApplPhysLett_2011}%
  \BibitemOpen
  \bibfield  {author} {\bibinfo {author} {\bibfnamefont {I.~A.~I.}\
  \bibnamefont {Al-Naib}}, \bibinfo {author} {\bibfnamefont {C.}~\bibnamefont
  {Jansen}}, \bibinfo {author} {\bibfnamefont {N.}~\bibnamefont {Born}}, \ and\
  \bibinfo {author} {\bibfnamefont {M.}~\bibnamefont {Koch}},\ }\href {\doibase
  10.1063/1.3562372} {\bibfield  {journal} {\bibinfo  {journal} {Appl. Phys.
  Lett.}\ }\textbf {\bibinfo {volume} {98}},\ \bibinfo {pages} {091107}
  (\bibinfo {year} {2011})}\BibitemShut {NoStop}%
\bibitem [{\citenamefont {Tuz}\ and\ \citenamefont
  {Prosvirnin}(2011)}]{Tuz_EurPhys_2011}%
  \BibitemOpen
  \bibfield  {author} {\bibinfo {author} {\bibfnamefont {V.~R.}\ \bibnamefont
  {Tuz}}\ and\ \bibinfo {author} {\bibfnamefont {S.~L.}\ \bibnamefont
  {Prosvirnin}},\ }\href {\doibase 10.1051/epjap/2011110145} {\bibfield
  {journal} {\bibinfo  {journal} {Eur. Phys. J. Appl. Phys.}\ }\textbf
  {\bibinfo {volume} {56}},\ \bibinfo {pages} {30401} (\bibinfo {year}
  {2011})}\BibitemShut {NoStop}%
\bibitem [{\citenamefont {Tuz}\ \emph {et~al.}(2012)\citenamefont {Tuz},
  \citenamefont {Butylkin},\ and\ \citenamefont {Prosvirnin}}]{Tuz_JOpt_2012}%
  \BibitemOpen
  \bibfield  {author} {\bibinfo {author} {\bibfnamefont {V.~R.}\ \bibnamefont
  {Tuz}}, \bibinfo {author} {\bibfnamefont {V.~S.}\ \bibnamefont {Butylkin}}, \
  and\ \bibinfo {author} {\bibfnamefont {S.~L.}\ \bibnamefont {Prosvirnin}},\
  }\href {http://stacks.iop.org/2040-8986/14/i=4/a=045102} {\bibfield
  {journal} {\bibinfo  {journal} {J. Opt.}\ }\textbf {\bibinfo {volume} {14}},\
  \bibinfo {pages} {045102} (\bibinfo {year} {2012})}\BibitemShut {NoStop}%
\bibitem [{\citenamefont {Meng}\ \emph {et~al.}(2012)\citenamefont {Meng},
  \citenamefont {Wu}, \citenamefont {Erni}, \citenamefont {Wu},\ and\
  \citenamefont {Lee}}]{Meng_MTT_2012}%
  \BibitemOpen
  \bibfield  {author} {\bibinfo {author} {\bibfnamefont {F.}~\bibnamefont
  {Meng}}, \bibinfo {author} {\bibfnamefont {Q.}~\bibnamefont {Wu}}, \bibinfo
  {author} {\bibfnamefont {D.}~\bibnamefont {Erni}}, \bibinfo {author}
  {\bibfnamefont {K.}~\bibnamefont {Wu}}, \ and\ \bibinfo {author}
  {\bibfnamefont {J.}~\bibnamefont {Lee}},\ }\href {\doibase
  10.1109/TMTT.2012.2209455} {\bibfield  {journal} {\bibinfo  {journal} {IEEE
  Trans. Microw. Theory Tech.}\ }\textbf {\bibinfo {volume} {60}},\ \bibinfo
  {pages} {3013} (\bibinfo {year} {2012})}\BibitemShut {NoStop}%
\bibitem [{\citenamefont {Tuong}\ \emph {et~al.}(2013)\citenamefont {Tuong},
  \citenamefont {Park}, \citenamefont {Rhee}, \citenamefont {Kim},
  \citenamefont {Jang}, \citenamefont {Cheong},\ and\ \citenamefont
  {Lee}}]{Tuong_ApplPhysLett_2013}%
  \BibitemOpen
  \bibfield  {author} {\bibinfo {author} {\bibfnamefont {P.~V.}\ \bibnamefont
  {Tuong}}, \bibinfo {author} {\bibfnamefont {J.~W.}\ \bibnamefont {Park}},
  \bibinfo {author} {\bibfnamefont {J.~Y.}\ \bibnamefont {Rhee}}, \bibinfo
  {author} {\bibfnamefont {K.~W.}\ \bibnamefont {Kim}}, \bibinfo {author}
  {\bibfnamefont {W.~H.}\ \bibnamefont {Jang}}, \bibinfo {author}
  {\bibfnamefont {H.}~\bibnamefont {Cheong}}, \ and\ \bibinfo {author}
  {\bibfnamefont {Y.~P.}\ \bibnamefont {Lee}},\ }\href {\doibase
  10.1063/1.4794173} {\bibfield  {journal} {\bibinfo  {journal} {Appl. Phys.
  Lett.}\ }\textbf {\bibinfo {volume} {102}},\ \bibinfo {pages} {081122}
  (\bibinfo {year} {2013})}\BibitemShut {NoStop}%
\bibitem [{\citenamefont {Yu}\ \emph {et~al.}(2013)\citenamefont {Yu},
  \citenamefont {Shi}, \citenamefont {Zhu}, \citenamefont {Liu},\ and\
  \citenamefont {Guan}}]{Yu_JOpt_2013}%
  \BibitemOpen
  \bibfield  {author} {\bibinfo {author} {\bibfnamefont {S.~W.}\ \bibnamefont
  {Yu}}, \bibinfo {author} {\bibfnamefont {J.~H.}\ \bibnamefont {Shi}},
  \bibinfo {author} {\bibfnamefont {Z.}~\bibnamefont {Zhu}}, \bibinfo {author}
  {\bibfnamefont {R.}~\bibnamefont {Liu}}, \ and\ \bibinfo {author}
  {\bibfnamefont {C.~Y.}\ \bibnamefont {Guan}},\ }\href
  {http://stacks.iop.org/2040-8986/15/i=7/a=075103} {\bibfield  {journal}
  {\bibinfo  {journal} {J. Opt.}\ }\textbf {\bibinfo {volume} {15}},\ \bibinfo
  {pages} {075103} (\bibinfo {year} {2013})}\BibitemShut {NoStop}%
\bibitem [{\citenamefont {Zhang}\ \emph
  {et~al.}(2013{\natexlab{b}})\citenamefont {Zhang}, \citenamefont {Zhao},
  \citenamefont {Zhou},\ and\ \citenamefont
  {Wang}}]{ZhangFuli_OptExpress_2013}%
  \BibitemOpen
  \bibfield  {author} {\bibinfo {author} {\bibfnamefont {F.}~\bibnamefont
  {Zhang}}, \bibinfo {author} {\bibfnamefont {Q.}~\bibnamefont {Zhao}},
  \bibinfo {author} {\bibfnamefont {J.}~\bibnamefont {Zhou}}, \ and\ \bibinfo
  {author} {\bibfnamefont {S.}~\bibnamefont {Wang}},\ }\href {\doibase
  10.1364/OE.21.019675} {\bibfield  {journal} {\bibinfo  {journal} {Opt.
  Express}\ }\textbf {\bibinfo {volume} {21}},\ \bibinfo {pages} {19675}
  (\bibinfo {year} {2013}{\natexlab{b}})}\BibitemShut {NoStop}%
\bibitem [{\citenamefont {Jain}\ \emph {et~al.}(2015)\citenamefont {Jain},
  \citenamefont {Moitra}, \citenamefont {Koschny}, \citenamefont {Valentine},\
  and\ \citenamefont {Soukoulis}}]{jain_advoptmater_2015}%
  \BibitemOpen
  \bibfield  {author} {\bibinfo {author} {\bibfnamefont {A.}~\bibnamefont
  {Jain}}, \bibinfo {author} {\bibfnamefont {P.}~\bibnamefont {Moitra}},
  \bibinfo {author} {\bibfnamefont {T.}~\bibnamefont {Koschny}}, \bibinfo
  {author} {\bibfnamefont {J.}~\bibnamefont {Valentine}}, \ and\ \bibinfo
  {author} {\bibfnamefont {C.~M.}\ \bibnamefont {Soukoulis}},\ }\href {\doibase
  10.1002/adom.201500222} {\bibfield  {journal} {\bibinfo  {journal} {Adv. Opt.
  Mater.}\ }\textbf {\bibinfo {volume} {3}},\ \bibinfo {pages} {1431} (\bibinfo
  {year} {2015})}\BibitemShut {NoStop}%
\bibitem [{\citenamefont {Sayanskiy}\ \emph {et~al.}(2018)\citenamefont
  {Sayanskiy}, \citenamefont {Danaeifar}, \citenamefont {Kapitanova},\ and\
  \citenamefont {Miroshnichenko}}]{Kapitanova_AdvOptMat}%
  \BibitemOpen
  \bibfield  {author} {\bibinfo {author} {\bibfnamefont {A.}~\bibnamefont
  {Sayanskiy}}, \bibinfo {author} {\bibfnamefont {M.}~\bibnamefont
  {Danaeifar}}, \bibinfo {author} {\bibfnamefont {P.}~\bibnamefont
  {Kapitanova}}, \ and\ \bibinfo {author} {\bibfnamefont {A.~E.}\ \bibnamefont
  {Miroshnichenko}},\ }\href {\doibase 10.1002/adom.201800302} {\bibfield
  {journal} {\bibinfo  {journal} {Adv. Opt. Mat.}\ }\textbf {\bibinfo {volume}
  {6}},\ \bibinfo {pages} {1800302} (\bibinfo {year} {2018})}\BibitemShut
  {NoStop}%
\bibitem [{\citenamefont {Lim}\ \emph {et~al.}(2010)\citenamefont {Lim},
  \citenamefont {Hong}, \citenamefont {Chen}, \citenamefont {Han},
  \citenamefont {Luk'yanchuk},\ and\ \citenamefont
  {Chong}}]{Lim_OptExpress_2010}%
  \BibitemOpen
  \bibfield  {author} {\bibinfo {author} {\bibfnamefont {C.~S.}\ \bibnamefont
  {Lim}}, \bibinfo {author} {\bibfnamefont {M.~H.}\ \bibnamefont {Hong}},
  \bibinfo {author} {\bibfnamefont {Z.~C.}\ \bibnamefont {Chen}}, \bibinfo
  {author} {\bibfnamefont {N.~R.}\ \bibnamefont {Han}}, \bibinfo {author}
  {\bibfnamefont {B.}~\bibnamefont {Luk'yanchuk}}, \ and\ \bibinfo {author}
  {\bibfnamefont {T.~C.}\ \bibnamefont {Chong}},\ }\href {\doibase
  10.1364/OE.18.012421} {\bibfield  {journal} {\bibinfo  {journal} {Opt.
  Express}\ }\textbf {\bibinfo {volume} {18}},\ \bibinfo {pages} {12421}
  (\bibinfo {year} {2010})}\BibitemShut {NoStop}%
\bibitem [{\citenamefont {Al-Naib}\ \emph {et~al.}(2012)\citenamefont
  {Al-Naib}, \citenamefont {Singh}, \citenamefont {Rockstuhl}, \citenamefont
  {Lederer}, \citenamefont {Delprat}, \citenamefont {Rocheleau}, \citenamefont
  {Chaker}, \citenamefont {Ozaki},\ and\ \citenamefont
  {Morandotti}}]{AlNaib_ApplPhysLett_2012}%
  \BibitemOpen
  \bibfield  {author} {\bibinfo {author} {\bibfnamefont {I.}~\bibnamefont
  {Al-Naib}}, \bibinfo {author} {\bibfnamefont {R.}~\bibnamefont {Singh}},
  \bibinfo {author} {\bibfnamefont {C.}~\bibnamefont {Rockstuhl}}, \bibinfo
  {author} {\bibfnamefont {F.}~\bibnamefont {Lederer}}, \bibinfo {author}
  {\bibfnamefont {S.}~\bibnamefont {Delprat}}, \bibinfo {author} {\bibfnamefont
  {D.}~\bibnamefont {Rocheleau}}, \bibinfo {author} {\bibfnamefont
  {M.}~\bibnamefont {Chaker}}, \bibinfo {author} {\bibfnamefont
  {T.}~\bibnamefont {Ozaki}}, \ and\ \bibinfo {author} {\bibfnamefont
  {R.}~\bibnamefont {Morandotti}},\ }\href {\doibase 10.1063/1.4745790}
  {\bibfield  {journal} {\bibinfo  {journal} {Appl. Phys. Lett.}\ }\textbf
  {\bibinfo {volume} {101}},\ \bibinfo {pages} {071108} (\bibinfo {year}
  {2012})}\BibitemShut {NoStop}%
\bibitem [{\citenamefont {Born}\ \emph {et~al.}(2014)\citenamefont {Born},
  \citenamefont {Al-Naib}, \citenamefont {Scheller}, \citenamefont {Jansen},
  \citenamefont {Moloney},\ and\ \citenamefont {Koch}}]{AlNaib_Conf_2014}%
  \BibitemOpen
  \bibfield  {author} {\bibinfo {author} {\bibfnamefont {N.}~\bibnamefont
  {Born}}, \bibinfo {author} {\bibfnamefont {I.}~\bibnamefont {Al-Naib}},
  \bibinfo {author} {\bibfnamefont {M.}~\bibnamefont {Scheller}}, \bibinfo
  {author} {\bibfnamefont {C.}~\bibnamefont {Jansen}}, \bibinfo {author}
  {\bibfnamefont {J.~V.}\ \bibnamefont {Moloney}}, \ and\ \bibinfo {author}
  {\bibfnamefont {M.}~\bibnamefont {Koch}},\ }in\ \href {\doibase
  10.1109/IRMMW-THz.2014.6956187} {\emph {\bibinfo {booktitle} {39th
  International Conference on Infrared, Millimeter, and Terahertz Waves
  (IRMMW-THz)}}}\ (\bibinfo {year} {2014})\ pp.\ \bibinfo {pages}
  {1--2}\BibitemShut {NoStop}%
\bibitem [{\citenamefont {Wu}\ \emph {et~al.}(2014)\citenamefont {Wu},
  \citenamefont {Yang}, \citenamefont {Zhao}, \citenamefont {Zheng},
  \citenamefont {Duan},\ and\ \citenamefont {Yuan}}]{Wu_OptExpress_2014}%
  \BibitemOpen
  \bibfield  {author} {\bibinfo {author} {\bibfnamefont {L.}~\bibnamefont
  {Wu}}, \bibinfo {author} {\bibfnamefont {Z.}~\bibnamefont {Yang}}, \bibinfo
  {author} {\bibfnamefont {M.}~\bibnamefont {Zhao}}, \bibinfo {author}
  {\bibfnamefont {Y.}~\bibnamefont {Zheng}}, \bibinfo {author} {\bibfnamefont
  {J.}~\bibnamefont {Duan}}, \ and\ \bibinfo {author} {\bibfnamefont
  {X.}~\bibnamefont {Yuan}},\ }\href {\doibase 10.1364/OE.22.014588} {\bibfield
   {journal} {\bibinfo  {journal} {Opt. Express}\ }\textbf {\bibinfo {volume}
  {22}},\ \bibinfo {pages} {14588} (\bibinfo {year} {2014})}\BibitemShut
  {NoStop}%
\bibitem [{\citenamefont {Tuz}\ \emph {et~al.}(2018{\natexlab{b}})\citenamefont
  {Tuz}, \citenamefont {Khardikov},\ and\ \citenamefont
  {Kivshar}}]{tuz_ACSPhotonics_2018}%
  \BibitemOpen
  \bibfield  {author} {\bibinfo {author} {\bibfnamefont {V.~R.}\ \bibnamefont
  {Tuz}}, \bibinfo {author} {\bibfnamefont {V.~V.}\ \bibnamefont {Khardikov}},
  \ and\ \bibinfo {author} {\bibfnamefont {Y.~S.}\ \bibnamefont {Kivshar}},\
  }\href {\doibase 10.1021/acsphotonics.8b00098} {\bibfield  {journal}
  {\bibinfo  {journal} {ACS Photonics}\ }\textbf {\bibinfo {volume} {5}},\
  \bibinfo {pages} {1871} (\bibinfo {year} {2018}{\natexlab{b}})}\BibitemShut
  {NoStop}%
\bibitem [{\citenamefont {Dmitriev}(2011)}]{Dmitriev_Metamat_2011}%
  \BibitemOpen
  \bibfield  {author} {\bibinfo {author} {\bibfnamefont {V.}~\bibnamefont
  {Dmitriev}},\ }\href {\doibase https://doi.org/10.1016/j.metmat.2011.04.003}
  {\bibfield  {journal} {\bibinfo  {journal} {Metamaterials}\ }\textbf
  {\bibinfo {volume} {5}},\ \bibinfo {pages} {14} (\bibinfo {year}
  {2011})}\BibitemShut {NoStop}%
\bibitem [{\citenamefont {Dmitriev}(2013)}]{Dmitriev_IEEEAntennas_2013}%
  \BibitemOpen
  \bibfield  {author} {\bibinfo {author} {\bibfnamefont {V.}~\bibnamefont
  {Dmitriev}},\ }\href {\doibase 10.1109/TAP.2012.2220316} {\bibfield
  {journal} {\bibinfo  {journal} {IEEE Trans. Antennas Propag.}\ }\textbf
  {\bibinfo {volume} {61}},\ \bibinfo {pages} {185} (\bibinfo {year}
  {2013})}\BibitemShut {NoStop}%
\bibitem [{com()}]{comsol}%
  \BibitemOpen
  \href
  {https://www.comsol.com/model/frequency-selective-surface-periodic-complementary-split-ring-resonator-15711}
  {\enquote {\bibinfo {title} {Frequency selective surface, periodic
  complementary split ring resonator},}\ }\bibinfo {note} {Comsol Application
  Gallery {ID:~15711}}\BibitemShut {NoStop}%
\bibitem [{\citenamefont {Barybin}\ and\ \citenamefont
  {Dmitriev}(2002)}]{barybin2002modern}%
  \BibitemOpen
  \bibfield  {author} {\bibinfo {author} {\bibfnamefont {A.~A.}\ \bibnamefont
  {Barybin}}\ and\ \bibinfo {author} {\bibfnamefont {V.~A.}\ \bibnamefont
  {Dmitriev}},\ }\href@noop {} {\emph {\bibinfo {title} {Modern Electrodynamics
  and Coupled-Mode Theory: Application to Guided-Wave Optics}}}\ (\bibinfo
  {publisher} {Rinton Press Princeton, New Jersey},\ \bibinfo {year}
  {2002})\BibitemShut {NoStop}%
\bibitem [{Sup()}]{Suppl_Mat}%
  \BibitemOpen
  \href@noop {} {}\bibinfo {note} {See Supplemental Material at (url) for
  animation representing the dynamic of the electromagnetic near-field within
  the super-cell of the all-dielectric metasurface. They correspond to those
  presented in Fig.~\ref{fig:simulated} of the manuscript.}\BibitemShut {Stop}%
\bibitem [{\citenamefont {Decker}\ \emph
  {et~al.}(2009{\natexlab{a}})\citenamefont {Decker}, \citenamefont {Burger},
  \citenamefont {Linden},\ and\ \citenamefont
  {Wegener}}]{Wegener_PhysRevB_2009}%
  \BibitemOpen
  \bibfield  {author} {\bibinfo {author} {\bibfnamefont {M.}~\bibnamefont
  {Decker}}, \bibinfo {author} {\bibfnamefont {S.}~\bibnamefont {Burger}},
  \bibinfo {author} {\bibfnamefont {S.}~\bibnamefont {Linden}}, \ and\ \bibinfo
  {author} {\bibfnamefont {M.}~\bibnamefont {Wegener}},\ }\href {\doibase
  10.1103/PhysRevB.80.193102} {\bibfield  {journal} {\bibinfo  {journal} {Phys.
  Rev. B}\ }\textbf {\bibinfo {volume} {80}},\ \bibinfo {pages} {193102}
  (\bibinfo {year} {2009}{\natexlab{a}})}\BibitemShut {NoStop}%
\bibitem [{\citenamefont {Mackay}(1989)}]{Mackay_ElectronLett_1989}%
  \BibitemOpen
  \bibfield  {author} {\bibinfo {author} {\bibfnamefont {A.}~\bibnamefont
  {Mackay}},\ }\href {\doibase 10.1049/el:19891088} {\bibfield  {journal}
  {\bibinfo  {journal} {Electron. Lett.}\ }\textbf {\bibinfo {volume} {25}},\
  \bibinfo {pages} {1624} (\bibinfo {year} {1989})}\BibitemShut {NoStop}%
\bibitem [{\citenamefont {Decker}\ \emph
  {et~al.}(2009{\natexlab{b}})\citenamefont {Decker}, \citenamefont {Linden},\
  and\ \citenamefont {Wegener}}]{Decker_OptLett_2009}%
  \BibitemOpen
  \bibfield  {author} {\bibinfo {author} {\bibfnamefont {M.}~\bibnamefont
  {Decker}}, \bibinfo {author} {\bibfnamefont {S.}~\bibnamefont {Linden}}, \
  and\ \bibinfo {author} {\bibfnamefont {M.}~\bibnamefont {Wegener}},\ }\href
  {\doibase 10.1364/OL.34.001579} {\bibfield  {journal} {\bibinfo  {journal}
  {Opt. Lett.}\ }\textbf {\bibinfo {volume} {34}},\ \bibinfo {pages} {1579}
  (\bibinfo {year} {2009}{\natexlab{b}})}\BibitemShut {NoStop}%
\bibitem [{\citenamefont {Johnson}\ \emph {et~al.}(1973)\citenamefont
  {Johnson}, \citenamefont {Ecker},\ and\ \citenamefont
  {Hollis}}]{johnson_IEEE_1973}%
  \BibitemOpen
  \bibfield  {author} {\bibinfo {author} {\bibfnamefont {R.~C.}\ \bibnamefont
  {Johnson}}, \bibinfo {author} {\bibfnamefont {H.~A.}\ \bibnamefont {Ecker}},
  \ and\ \bibinfo {author} {\bibfnamefont {J.~S.}\ \bibnamefont {Hollis}},\
  }\href {\doibase 10.1109/PROC.1973.9358} {\bibfield  {journal} {\bibinfo
  {journal} {Proc. IEEE}\ }\textbf {\bibinfo {volume} {61}},\ \bibinfo {pages}
  {1668} (\bibinfo {year} {1973})}\BibitemShut {NoStop}%
\end{thebibliography}%

\end{document}